\documentclass[preprint,english,superscriptaddress,graphicx]{revtex4-1}
\usepackage[latin9]{inputenc}
\usepackage{babel}
\usepackage{amsmath}
\usepackage{amssymb}
\usepackage{graphicx}
\usepackage{color}
\usepackage{esint}
\usepackage[unicode=true,pdfusetitle,
 bookmarks=true,bookmarksnumbered=false,bookmarksopen=false,
 breaklinks=false,pdfborder={0 0 1},backref=false,colorlinks=false]
 {hyperref}
\usepackage{breakurl}

\makeatletter
\@ifundefined{textcolor}{}
{%
 \definecolor{BLACK}{gray}{0}
 \definecolor{WHITE}{gray}{1}
 \definecolor{RED}{rgb}{1,0,0}
 \definecolor{GREEN}{rgb}{0,1,0}
 \definecolor{BLUE}{rgb}{0,0,1}
 \definecolor{CYAN}{cmyk}{1,0,0,0}
 \definecolor{MAGENTA}{cmyk}{0,1,0,0}
 \definecolor{YELLOW}{cmyk}{0,0,1,0}
}

\usepackage{amsfonts}\usepackage{comment}
\newcommand{\ud}{\,\mathrm{d}}
\newcommand{\pderiv}[2]{\frac{\partial #1}{\partial #2}} 

\makeatother

\begin{document}

\title{Geodesic acoustic modes in a fluid model of tokamak plasma : the
effects of finite beta and collisionality}

\author{Rameswar Singh}
\affiliation{Laboratoire de Physique des Plasmas, Ecole Polytechnique, 91128 Palaiseau Cedex, France}
\email{rameswar.singh@lpp.polytechnique.fr}
\author{A Storelli}
\affiliation{Laboratoire de Physique des Plasmas, Ecole Polytechnique, 91128 Palaiseau Cedex, France}
\author{Ö D Gürcan}
\affiliation{Laboratoire de Physique des Plasmas, Ecole Polytechnique, 91128 Palaiseau Cedex, France}
\author{P Hennequin}
\affiliation{Laboratoire de Physique des Plasmas, Ecole Polytechnique, 91128 Palaiseau Cedex, France}
\author{L Vermare}
\affiliation{Laboratoire de Physique des Plasmas, Ecole Polytechnique, 91128 Palaiseau Cedex, France}
\author{P Morel}
\affiliation{Laboratoire de Physique des Plasmas, Ecole Polytechnique, 91128 Palaiseau Cedex, France}
\affiliation{Université Paris-Sud, UMR 7648, Laboratoire de Physique des Plasmas, 91128 Palaiseau, France}

\author{R Singh}
\affiliation{WCI Center for Fusion Theory, National Fusion Research Institute, Daejeon 305-333, South Korea}
\affiliation{Institute for Plasma Research, Bhat, Gandhinagar - 382 428, India}

\date{\today}
\begin{abstract}
Starting from the Braginskii equations, relevant for the tokamak edge
region, a complete set of nonlinear equations for the geodesic acoustic
modes (GAM) has been derived which includes collisionality, plasma
beta and external sources of particle, momentum and heat. Local linear
analysis shows that the GAM frequency increases with collisionality at 
low radial wave number $k_{r}$ and decreases at high $k_{r}$. GAM frequency also 
decreases with plasma beta. Radial profiles of GAM frequency for two
Tore Supra shots, which were part of a collisionality scan, are compared
with these calculations. Discrepency between experiment and theory
is observed, which seems to be explained by a finite $k_{r}$ for
the GAM when flux surface averaged density $\langle n \rangle$ and
temperature $\langle T \rangle$ are assumed to 
vanish. It is shown that this agreement is incidental and
self-consistent inclusion of $\langle n \rangle$ and
$\langle T \rangle$ responses enhances the disagreement more with $k_r$ at 
high $k_{r}$ . 
So the discrepancy
between the linear GAM calculation, (which persist also for more ``complete''
linear models such as gyrokinetics) can probably not be resolved by
simply adding a finite $k_{r}$. 
\end{abstract}
\maketitle

\section{Introduction}

Common wisdom in fusion plasma science is that the transport of heat
and particles in tokamaks are largely due to micro-turbulence driven
by background gradients of density, temperature, momentum etc. The
turbulence saturates via mode coupling, and in particular by interactions
with self-generated large scale flow structures such as zonal flows\citep{diamond},and
in some cases, especially near the edge, with geodesic acoustic modes
(GAMs)\citep{winsor:68,miki:2007,miki:2008}. GAMs are an important
class of oscillating zonal flows that appear due to toroidal geometry
(i.e. due to geodesic curvature), and are easily observable in tokamak
experiments due to their finite frequency. They are usually classified
as an $m=n=0$ perturbation in potential coupled with an $m=1$, $n=0$
perturbation in density or pressure, where $m$ and $n$ are the poloidal
and the toroidal mode numbers respectively. GAMs are linearly
damped unless fast particles are present\citep{fu:2008,fu:2011,qiu:2010,qiu:2012,zarzoso:2012,wang:2014,girardo:2014}.
Otherwise they are excited by nonlinear processes like turbulent reynolds
stresses\citep{nikhil:2007,nikhil:2008,guzdar:2008,zonca:epl2008,guzdar:2009,yu:2013,chen:2014},
poloidally asymmetric particle fluxes\citep{itoh:2005} and heat fluxes\citep{hallatschek:2001}.
\foreignlanguage{american}{Due to their finite frequency (usually
a few $kHz$), distinct from that of broadband turbulence, GAMs are
easier to detect, and thus have been observed on several tokamaks
such as ASDEX Upgrade (AUG) \citep{conway:05} using Doppler backscattering
(DBS), TEXTOR \citep{flecken:2006}using O-mode correlation reflectometer,
and DIIID \citep{mckee:06} using beam emission spectroscopy (BES).
As of today, GAMS are observed in the majority of the tokamaks in
the world including recent observations of GAMs in Tore Supra \citep{vermare:12}
using a DBS system. The common aspect of these measurements is that
the GAMs are most prominent in the edge region, right inside the last
closed flux surface, and extend into the near edge region (sometimes
called the no man's land due to a seemingly systematic discrepency
between simulation and experiment \citep{holland:09}). While gyrokinetics
is accepted widely as the most general formulation for strongly magnetised
plasmas of tokamak fusion devices, the applicability of gyrokinetic
vs. fluids models is still somewhat open to debate in the edge region.
For most existing tokamaks, as one goes from the core to the edge,
the collisions start to play a role, and the parallel connection length
increases (since the safety factor $q$ increases), dissipative drift
waves, or resistive balloning modes, start to become important, therefore
the validity of a fluid description including the effects of collisions
may actually be justified\citep{rafiq:10,militello:11}.}

In this spirit, here we will develop a simple two fluid model for
the description of the GAM, using Braginskii equations \citep{braginskii:65,zeiler:97} within
a drift expansion, in order to include the effects of collisions.
In particular we include equations of contunity, momentum and heat
for ions and electrons (i.e. using a generalized Ohm's law for electrons)
coupled with the Ampère's law. The formulation allows us to include
$v_{\parallel}$, $A_{\parallel}$, $T_{i}$ and $T_{e}$ perturbations
of the GAM in a full set of nonlinear equations, which can be linearized
and solved to obtain the GAM frequency including the effects of collisions
and finite $\beta$, and finite radial mode number $k_{r}$.

The computed frequency is then compared with the radial profile of
GAM frequency that is observed in Tore Supra during a collisionality
scan (assuming $k_{r}\approx0$). There is an apparent, systematic
discrepancy between the theory and the experiment, which seems to
be explained when a finite $k_{r}$ is introduced for the GAM calculation 
assuming 
flux surface averaged density $\langle n \rangle$, ion temperature $\langle T_{i}\rangle$
and flux surface averaged electron temperature $\langle T_{e} \rangle$ to be zero. 
However we believe that this agreement is incidental since it breaks
down when higher harmonics ($m=2$ etc.) are included in the calculation\citep{nguyen:2008,rameswar1:2014}.
The agreement also breaks down on self consistent inclusion of $\langle n \rangle$,
$\langle T_{i,e} \rangle$ responses on  the GAM dispersion for $m=1$.
This indicates that the discrepancy between the linear GAM calculation
and the experiment, which persist also for more ``complete'' linear
models such as gyrokinetics, is probably significant and can not be
resolved by simply adding a finite $k_{r}$. 

The remainder of this paper is organized as follows. The complete
set of nonlinear electromagnetic equations with collisionality are
obtained in Section\ref{sec:NONLINEAR-MODEL-EQUATIONS} from the drift
reduction of Braginskii equations. The fully nonlinear equations for
GAMs are obtained in Section\ref{sec:Geodesic-acoustic-mode} by taking
appropriate flux surface averagings of the drift reduced electron
and ion equations. Linear GAM dispersion properties are obtained in
Section\ref{sub:Linear-GAMs-dispersion} and comparison with experimental
data are presented in Section\ref{sub:Comparison-with-experiment}.
Finally the paper is concluded in Section\ref{sec:Conclusions}.

\section{NONLINEAR MODEL EQUATIONS\label{sec:NONLINEAR-MODEL-EQUATIONS}}

In order to formulate the nonlinear theory of electromagnetic geodesic
acoustic modes (GAMs), we start with the simple two fluid Braginskii
equations\citep{braginskii:65,zeiler:97}, where we keep the following:
(i) the non adiabatic electron response with $\delta T_{e}\ne0$ ,
and the electron-ion collisionality $\nu_{ei}$. The model equations
for GAMs are then derived from the density, momentum and temperature
equations for each species $j(=i,e)$. 
\begin{equation}
\frac{\partial n_{j}}{\partial t}+\vec{\nabla}\cdot(n_{j}\vec{v}_{j})=0\label{GrindEQ__1_}
\end{equation}
 
\begin{equation}
m_{i}n_{i}\left(\frac{\partial v_{i}}{\partial t}+\vec{v}_{i}\cdot\vec{\nabla}\right)v_{i}=-\vec{\nabla}p_{i}-\vec{\nabla}\cdot\bar{\bar{\pi_{i}}}+e\left(\vec{E}+\frac{1}{c}\vec{v}_{i}\times\vec{B}\right)+\vec{R}_{ie}\label{GrindEQ__3_}
\end{equation}

\begin{equation}
0=-\vec{\nabla}p_{e}-e\left(\vec{E}+\frac{1}{c}\vec{v}_{e}\times\vec{B}\right)+\vec{R}_{ei}\label{GrindEQ__3_}
\end{equation}

\begin{equation}
\frac{3}{2}n_{j}\left(\frac{\partial}{\partial t}+\vec{v}_{j}\cdot\vec{\nabla}\right)T_{j}+p_{j}\,\vec{\nabla}\cdot\vec{v}_{j}=-\vec{\nabla}\cdot\vec{q}_{j}\label{GrindEQ__4_}
\end{equation}
 where 
\begin{eqnarray}
q_{j}=0.71nT_{j}\vec{U}_{||}-\kappa_{||}\nabla_{||}T_{j}+\kappa_{\perp}\nabla_{\perp}T_{j}+q_{*j}+\frac{3}{2}\nu_{j}\frac{nT_{j}}{\omega_{cj}}\hat{b}\times\vec{U}
\end{eqnarray}
 Here the mass of the electron is neglected and $\vec{E}=\vec{\nabla}\phi-(1/c)\partial{\vec{A}}/\partial{t}$.
The collisional momentum transfer term is given by $\vec{R}_{ei}=-\vec{R}_{ie}=ne\eta_{||}\vec{J}_{||}-0.71n\vec{\nabla}_{||}T_{e}$,
where $J_{||}=en(v_{i}-v_{||e})$, and $\vec{U}$ is the relative
velocity between species $j$ and $i$. The thermal conductivities
for electrons are given by $\kappa_{||e}=3.16n_{e}T_{e}/m_{e}\nu_{e}$,
$\kappa_{\perp e}=4.66n_{e}T_{e}\nu_{e}/m_{e}\omega_{ce}^{2}$ and
for ions $\kappa_{||i}=3.9n_{i}T_{i}/m_{i}\nu_{i}$, $\kappa_{\perp i}=2n_{i}T_{i}\nu_{i}/m_{i}\omega_{ci}^{2}$.
The diamagnetic heat flux is taken as $q_{*j}=\frac{5}{2}\frac{p_{j}}{m_{j}\omega_{cj}}\hat{b}\times\vec{\nabla}T_{j}$.

In order to develop a drift expansion, we consider ion and electron
perpendicular drift velocities in the low frequency regime ($\omega<<\omega_{ci}$,
$\nu_{ei}<<\omega_{ci}$; $\omega$, $\omega_{ci}=eB/m_{i}c$ mode
frequency, ion cyclotron frequency, respectively). These drift velocities
consist of the$\vec{E}\times\vec{B}$ drift, the ion and the electron
diamagnetic drifts, the ion polarization drift: 
\begin{eqnarray}
\vec{v}_{E}=(c/B^{2})\vec{B}\times\vec{\nabla}\delta\phi\\
\vec{v}_{*pi}=(c/en_{i}B^{2})\vec{B}\times\vec{\nabla}\delta p_{i}\\
\vec{v}_{*pe}=-(c/en_{e}B^{2})\vec{B}\times\vec{\nabla}\delta p_{e}\\
\vec{v}_{pi}=-\frac{c}{B\omega_{ci}}\left(\frac{\partial}{\partial t}+(\vec{v}_{E}+\vec{v}_{*pi})\cdot\vec{\nabla}\right)\vec{\nabla}_{\bot}\delta\phi
\end{eqnarray}

\noindent For the equilibrium scale lengths that are larger than the
perturbation scales (i.e.,$k_{\bot}L<1$), we can separate the equilibrium
($f_{0}$) and the fluctuating parts ($\delta f$) in the above set
of equations as $f=f_{0}+\delta f$. The complete set of resulting
reduced nonlinear equations for the perturbations ($\delta f$) are
provided in the next subsection which is written in the following
normalization scheme. The space time scales are normalized as $r=r/\rho_{s}$,
$\nabla_{\parallel}\equiv L_{n}\nabla_{\parallel}$, $t=tc_{s}/{L_{n}}$.
The field quantities are normalized to their mixing length levels:
$\phi=({e\delta\phi}/{T_{e}})({L_{n}}/{\rho_{s}})$, $n_{i}=({\delta n_{i}}/n_{0})({L_{n}}/{\rho_{s}})$,
$v=({\delta v_{\parallel i}}/{c_{s}})({L_{n}}/{\rho_{s}})$, $p_{i}=({\delta p_{i}}/{P_{e0}})({L_{n}}/{\rho_{s}})$,
$A_{||}=(2L_{n}c_{s}/\beta\rho_{s}c)(e\delta A_{||}/T_{e0})$. The
remaining dimensionless parameters are : $\eta_{i}=L_{n_{i0}}/L_{T_{i0}}$,
$K=\tau_{i}(1+\eta_{i})$, $\tau_{i}=T_{i0}/T_{e0}$, $\beta=8\pi P_{0e}/B_{0}^{2}$,
$L_{f}=-\ud lnf/\ud x$, $\eta_{e}=L_{n_{e0}}/L_{T_{e0}}$, $\nu=0.51m_{e}\nu_{ei}L_{n}/m_{i}c_{s}$.
The electron ion collision frequency is calculated from $\nu_{ei}=nZ^{2}ln\lambda/(1.09\times10^{16}T_{e}^{3/2})$
where $ln\lambda=15.2-log(n/10^{20})+log(T_{e})$\citep{wesson:97}.
$\rho_{s}=c_{s}/\omega_{ci}$ is the ion sound radius. The nonlinearities
in the following equations originate mainly from the $E\times B$
drift nonlinearity i.e., $\vec{v}_{E\times B}\cdot\vec{\nabla}f=\left[\phi,f\right]$,
the polariztion drift nonlinearity $\vec{v}_{E\times B}\cdot\vec{\nabla}\nabla_{\perp}^{2}f=\left[\phi,\nabla_{\perp}^{2}f\right]$
and the nonlinearity due to the parallel gradients with fluctuating
magnetic fields, from $\nabla_{||}=\nabla_{||}^{0}+\tilde{\delta\vec{B}}_{\perp}\cdot\vec{\nabla}=\nabla_{||}^{0}-(\beta/2)\left[A_{||},\quad\right]$,
$\nabla_{||}^{0}$ being derivative along the equilibrium magnetic
field. The $\beta$ effects enter via perpendicular magnetic field
line bending effect through the expressions for $E_{\parallel}$ and
the instantaneous parallel derivative.

\noindent Such models has also been used for edge turbulence simulations
in the references\citep{ricci:2010,beyer:1998,naulin:2005,bdscott:2005,Dudson:2009}

\subsection{Electron response}

When the drift expansion is considered in toroidal geometry, the electron
continuity equation for density perturbation takes the form: 
\begin{eqnarray}
 &  & \pderiv{n_{e}}{t}+\frac{1}{r}\pderiv{\phi}{\theta}-\varepsilon_{n}(\cos\theta\frac{1}{r}\frac{\partial}{\partial\theta}+\sin\theta\frac{\partial}{\partial r})(\phi-n_{e}-T_{e})-\nabla_{||}\left(J_{||}-v_{||}\right)\nonumber \\
 &  & =\left[\phi,n_{e}\right]-\frac{\beta}{2}\left[A_{||},J_{||}-v_{||}\right]
 \label{ern}
\end{eqnarray}

 These and the following equations has been derived assuming large
aspect ratio curcular flux surfaces. The second and third terms in
the above equation results from the $E\times B$ convection of equilibrium
density, the sum of divergence of $E\times B$ diamagnetic drifts
due to inhomogenous magnetic fields of the tokamak, respectively.
The first term on the right hand is the $E\times B$ convective nonlinearity
and the second term results from parallel derivative nonlinearity
due to perpendicular magnetic fluctuations. In the limit $\omega<<k_{||}c_{s}$
the perturbed parallel momentum equation for electrons reads: 
\begin{eqnarray}
\nu J_{||}=-\nabla_{||}\left(\phi-n_{e}-1.71T_{e}\right)-\frac{\beta}{2}\left[\pderiv{}{t}+\left(1+1.71\eta_{e}\right)\frac{1}{r}\pderiv{}{\theta}\right]A_{||}+\frac{\beta}{2}\left[A_{||},\phi-n_{e}-1.71T_{e}\right]
\end{eqnarray}
Similarly, the electron temperature perturbation equation, in the
same expansion, becomes: 
\begin{eqnarray}
 &  & \pderiv{}{t}(T_{e}-\frac{2}{3}n_{e})+\frac{5}{3}\varepsilon_{n}(\cos\theta\frac{1}{r}\frac{\partial}{\partial\theta}+\sin\theta\frac{\partial}{\partial r})T_{e}+\left(\eta_{e}-\frac{2}{3}\right)\frac{1}{r}\pderiv{\phi}{\theta}-\frac{1.07}{\nu}\nabla_{||}^{2}T_{e}\nonumber \\
 &  & =-\left[\phi,T_{e}-\frac{2}{3}n{e}\right]+\frac{1.07}{\nu}\left\lbrace \nabla_{||}\left(-\eta_{e}\frac{\beta}{2r}\pderiv{A_{||}}{\theta}-\frac{\beta}{2}\left[A_{||},T_{e}\right]\right)-\frac{\beta}{2}\left[A_{||},\nabla_{||}T_{e}\right.\right.\nonumber \\
 &  & \left.\left.-\eta_{e}\frac{\beta}{2r}\pderiv{A_{||}}{\theta}-\frac{\beta}{2}\left[A_{||},T_{e}\right]\right]\right\rbrace 
\label{ert}
 \end{eqnarray}


The first term on the right hand side is the $E\times B$ convective
nonlinearity and the second term results from $\tilde{\delta\vec{B}}_{\perp}$part
of the parallel derivative. This equation has been derived using the
continuity equation for $\nabla\cdot\vec{v}$ and the diamagnetic
heat flux cancelation.

Finally, the $J_{||}$ is related to the parallel vector potential
$A_{\parallel}$ via the Ampères law: 
\begin{eqnarray}
J_{||}=-\nabla_{\perp}^{2}A_{||}
\end{eqnarray}

\subsection{Ion response}

The equations for the ion dynamics can similarly be obtained from
the two fluid Braginskii equations using the drift expansion, with
the aforementioned assumptions. The resulting equation for the continuity
of ions take the form: 
\begin{eqnarray}
 &  & \frac{\partial n_{i}}{\partial t}+\frac{1}{r}\frac{\partial\phi}{\partial\theta}-\epsilon_{n}\left(\cos\theta\frac{1}{r}\pderiv{}{\theta}+\sin\theta\pderiv{}{r}\right)\left(\phi+\tau_{i}n_{i}+\tau_{i}T_{i}\right)-\left(\frac{\partial}{\partial t}-K\frac{1}{r}\pderiv{}{\theta}\right)\nabla_{\perp}^{2}\phi\nonumber \\
 &  & +\nabla_{||}^{0}v_{||}=-\left[\phi,n_{i}\right]+\vec{\nabla}\cdot\left[\phi+p_{i},\vec{\nabla}_{\perp}\phi\right]+\frac{\beta}{2}\left[A_{||},v_{||}\right]\label{ec}
\end{eqnarray}
Similar to the electron contunity equation, the second and third terms
correspond to the $E\times B$ convection of the bacground density
and the effects of inhomogenous magnetic field respectively, whilethe
fourth term comes from the divergence of the polarization drift. The
first term on the right hand side is the $E\times B$ convective nonlinearity,
the second term is the polarization nonlinearity, and the third term
results from parallel derivative nonlinearity due to perpendicular
magnetic fluctuations. Adding the electron momentum equation to the
ion momentum equation and then using the electron temperature equation,
one obtains the parallel ion velocity perturbation equation: 
\begin{eqnarray}
 &  & \pderiv{v_{||}}{t}-2\tau_{i}\varepsilon_{n}\left(\cos\theta\frac{1}{r}\pderiv{}{\theta}+\sin\theta\pderiv{}{r}\right)v_{||}+\nabla_{||}\left[\tau_{i}n_{i}+\tau_{i}T_{i}+n_{e}+T_{e}\right]\nonumber \\
 &  & -\frac{\beta}{2}\left(\tau_{i}(1+\eta_{i})+1+\eta_{e}\right)\frac{1}{r}\pderiv{A_{||}}{\theta}=-\left[\phi,v_{||}\right]+\frac{\beta}{2}\left[A_{||},p_{i}+p_{e}\right]+n_{i}\nabla_{||}\left(p_{i}+p_{e}\right)\label{iv}
\end{eqnarray}
Here, the first term on the right hand side is the $E\times B$ convective
nonlinearity, the second term results from perpendicular magnetic
fluctuation induced parallel derivative nonlinearity, and the third
term is the parallel acceleration term. 

Using the ion continuity equation for $\nabla\cdot\vec{v}$ and the
ion diamagnetic heat flux cancelation the ion temperature perturbation
equation becomes: 
\begin{eqnarray}
\frac{\partial}{\partial t}\left(T_{i}-\frac{2}{3}n_{i}\right)-\frac{5}{3}\varepsilon_{n}\left(\cos\theta\frac{1}{r}\pderiv{}{\theta}+\sin\theta\pderiv{}{r}\right)T_{i}+\left(\eta_{i}-\frac{2}{3}\right)\frac{1}{r}\pderiv{\phi}{\theta}=-\left[\phi,T_{i}-\frac{2}{3}n_{i}\right]\label{it}
\end{eqnarray}

Adding the electron and ion continuity equations after multiplying
by respective charges and then assuming quasineutrality for the perturbations
results in the plasma vorticity equation: 
\begin{eqnarray}
 &  & \left(\frac{\partial}{\partial t}-K\frac{1}{r}\pderiv{}{\theta}\right)\nabla_{\perp}^{2}\phi+\varepsilon_{n}\left(\cos\theta\frac{1}{r}\pderiv{}{\theta}+\sin\theta\pderiv{}{r}\right)\left[\tau_{i}n_{i}+\tau_{i}T_{i}+n_{e}+T_{e}\right]-\nabla_{||}^{0}J_{||}\label{vorticity}\\
 &  & =-\vec{\nabla}\cdot\left[\phi+p_{i},\vec{\nabla}_{\perp}\phi\right]+\frac{\beta}{2}\left[A_{||},\vec{\nabla}_{\perp}^{2}A_{||}\right]
\end{eqnarray}
where the $\beta$ dependent nonlinear term comes from the perpendicular
magnetic perturbations. 

The set of equations presented above for ions and electrons, provide
a full drift-Braginskii system that can be used to describe the GAM
oscillations including corrections due to finite $\beta$ and collisionality,
which may be relevant for the edge and near edge regions of tokamaks
where GAMs have traditionally been observed.

\section{Geodesic acoustic mode\label{sec:Geodesic-acoustic-mode}}

GAMs are low poloidal mode number ($m$) axisymmetric fluctuations
, that are supported by the geodesic component of the equilibrium
magnetic curvature in tokamaks. In order to derive the set of equations
that can be used to describe them, we start by taking the flux surface
average of the vorticity equation \eqref{vorticity}:

\begin{eqnarray}
 &  & \frac{\partial}{\partial t}\nabla_{r}^{2}\left\langle \phi\right\rangle +\varepsilon_{n}\pderiv{}{r}\left[\left(1+\tau_{i}\right)\left\langle n\sin\theta\right\rangle +\tau_{i}\left\langle T_{i}\sin\theta\right\rangle +\left\langle T_{e}\sin\theta\right\rangle \right]\nonumber \\
 &  & =-\left\langle \vec{\nabla}\cdot\left[\phi+p_{i},\vec{\nabla}_{\perp}\phi\right]\right\rangle +\frac{\beta}{2}\left\langle \left[A_{||},\vec{\nabla}_{\perp}^{2}A_{||}\right]\right\rangle \label{mvor}
\end{eqnarray}
 This shows that the flux surface averaged potential is linearly coupled
to the $m=1$ of density and temperature perturbations in the form
of flux surface averaged $\left\langle n\sin\theta\right\rangle $,
$\left\langle T_{i}\sin\theta\right\rangle $ and $\langle T_{e}\sin\theta\rangle$
perturbations. This coupling happens due to the \emph{geodesic curvature.}
The effect of normal curvature for $m=1$ mode is of the order of
$\rho_{s}/r\thicksim\rho_{s}/a\thicksim10^{-3}$ and hence can be
neglected. The nonlinear terms on the right hand side constitute the
flux surface averaged poloidal momentum flux/Reynolds stres and Maxwell
stress and act as turbulent source/sink of vorticity. Equation \eqref{mvor}
should be supplemented by the equation for $\left\langle n\sin\theta\right\rangle $,
which can be obtained by multiplying the ion continiuity equation
by $\sin\theta$ followed by flux surface averaging: 
\begin{eqnarray}
 &  & \pderiv{}{t}\left\langle n\sin\theta\right\rangle -\pderiv{}{t}\nabla_{r}^{2}\left\langle \phi\sin\theta\right\rangle -\frac{\varepsilon_{n}}{2}\nabla_{r}\left(\left\langle \phi\right\rangle +\tau_{i}\left\langle n\right\rangle +\tau_{i}\left\langle T_{i}\right\rangle \right)-\frac{\varepsilon_{n}}{2q}\left\langle v_{||}\cos\theta\right\rangle \nonumber \\
 &  & =-\left\langle \left[\phi,n\right]\sin\theta\right\rangle -\left\langle \vec{\nabla}\cdot\left[\phi+p_{i},\vec{\nabla}_{\perp}\phi\right]\sin\theta\right\rangle +\frac{\beta}{2}\left\langle \left[A_{||},v_{||}\right]\sin\theta\right\rangle +\left\langle S_{n}\sin\theta\right\rangle \label{zcs}
\end{eqnarray}

Here, the term representing $E\times B$ convection of equilibrium
density gradient is again dropped due to the fact that it is of the
order of $\rho_{s}/r\thicksim\rho_{s}/a\thicksim10^{-3}$. Equation
\eqref{zcs} shows a linear coupling, this time with $m=1$ of the
parallel velocity fluctution in the form of the flux surface averaged
quantity $\left\langle v_{||}\cos\theta\right\rangle .$ As is usually
the case for GAMs, it is assumed that $\langle(\phi,n,T_{i})\sin^{2}\theta\rangle=(\langle\phi\rangle,\langle n\rangle,\langle T_{i}\rangle)\langle\sin^{2}\theta\rangle$,
that is couplings to $m=2$ and higher harmonics are ignored. In fact,
the retention of $m=2$ demands for the equations for $m=3$ and so
on, which continues up to infinity. This closure problem, and its
possible resolutionwill be discussed in a future publication\citep{rameswar1:2014}.
The nonlinear terms on the right hand side acting as source/sink are
poloidally asymmetric turbulent particle flux, poloidal momentum flux
(i.e. Reynolds stress), and the electromagnetic component of parallel
momentum flux. An asymmetry in the external particle source may also
act as a source for the $\left\langle n\sin\theta\right\rangle $
component and therefore the GAM. The equation for $\left\langle v_{||}\cos\theta\right\rangle $
can be obtained by multiplying the parallel ion velocity equation
\eqref{iv} by $\cos\theta$ followed by flux surface averaging: 
\begin{eqnarray}
 &  & \frac{\partial}{\partial t}\left\langle v_{||}\cos\theta\right\rangle +\frac{\varepsilon_{n}}{2q}\left[\left(1+\tau_{i}\right)\left\langle n\sin\theta\right\rangle +\tau_{i}\left\langle T_{i}\sin\theta\right\rangle +\left\langle T_{e}\sin\theta\right\rangle \right]=-\left\langle \left[\phi,v_{||}\right]\cos\theta\right\rangle \nonumber \\
 &  & +\frac{\beta}{2}\left\langle \left[A_{||},p_{i}+p_{e}\right]\cos\theta\right\rangle +\left\langle \left(n\nabla_{||}\left(p_{i}+p_{e}\right)\right)\cos\theta\right\rangle +\left\langle S_{v}\cos\theta\right\rangle \label{gv}
\end{eqnarray}
where the normal curvature term and second harmonic terms like $\langle v_{\parallel}\sin2\theta\rangle$
are again dropped. The various nonlinear terms on right hand side
of the above equation can be identified as follows: The first term
coming from the $E\times B$ convective nonlinearity is the $\cos\theta$
weighted, flux surface averaged, divergence of the parallel velocity
flux. The second term is the electromagnetic analog due to perpendcular
magnetic perturbation. The third term is the flux surface average
of the turbulent parallel acceleration weighted by $\cos\theta$.
This term survives only when there is a $k_{\parallel}$symmetry breaking
mechanism present\citep{luwang:2013}, which breaks the dipolar structure
of acceleration in $\theta$. The last term is the $\theta$ symmetric
part of the external velocity/ momentum source. The equation for $\left\langle T_{i}\sin\theta\right\rangle $
, which appears in equations \eqref{mvor} and \eqref{gv} can be
obtained by multiplying the ion temperature perturbation equation
by $\sin\theta$ followed by flux surface averaging: 
\begin{eqnarray}
\frac{\partial}{\partial t}\left(\left\langle T_{i}\sin\theta\right\rangle -\frac{2}{3}\left\langle n\sin\theta\right\rangle \right)-\frac{5}{3}\frac{\varepsilon_{n}}{2}\nabla_{r}\left\langle T_{i}\right\rangle =-\left\langle \left[\phi,T_{i}-\frac{2}{3}n\right]\sin\theta\right\rangle +\left\langle S_{T}\sin\theta\right\rangle \label{zt}
\end{eqnarray}
 The first nonlinear term on the right hand side is the divergence
of flux surface average of the $\sin\theta$ weighted heat flux minus
$2/3$ times particle flux. The second term is the poloidally asymmetric
part of the external heating.

The electron temperature equation for $\left\langle T_{e}\sin\theta\right\rangle $
reads: 
\begin{eqnarray}
 &  & \frac{\partial}{\partial t}\left(\left\langle T_{e}\sin\theta\right\rangle -\frac{2}{3}\left\langle n\sin\theta\right\rangle \right)+\frac{5}{3}\frac{\varepsilon_{n}}{2}\nabla_{r}\left\langle T_{e}\right\rangle +\frac{1.07}{\nu}\left(\frac{\varepsilon_{n}}{2q}\right)^{2}\left\langle T_{e}\sin\theta\right\rangle -0.71\frac{\varepsilon_{n}}{2q}\nabla_{r}^{2}\left\langle A_{||}\cos\theta\right\rangle \nonumber \\
 &  & =-\left\langle \left[\phi,T_{i}-\frac{2}{3}n\right]\sin\theta\right\rangle +\left\langle S_{T_{e}}\sin\theta\right\rangle \label{zt}
\end{eqnarray}

The first nonlinear term on the right hand side is the divergence
of heat flux minus particle flux weighted by $\sin\theta$. The second
term is the poloidally asymmetric part of external heating


Multiplying parallel electron velocity equation by $\cos\theta$ and
then taking the flux surface average gives: 
\begin{eqnarray}
 &  & \left(\nu\nabla_{r}^{2}-\frac{\beta}{2}\pderiv{}{t}\right)\left\langle A_{||}\cos\theta\right\rangle -\frac{\varepsilon_{n}}{2q}\left[\left\langle \phi\sin\theta\right\rangle -\left\langle n\sin\theta\right\rangle -1.71\left\langle T_{e}\sin\theta\right\rangle \right]\nonumber \\
 &  & =-\frac{\beta}{2}\left\langle \left[A_{||},\phi-n-1.71T_{e}\right]\cos\theta\right\rangle \label{mev}
\end{eqnarray}

The electromagnetic nonlinear term on the right hand side comes from
the perpendicular magnetic perturbation from the parallel derivative.
Finally the equation for $\left\langle \phi\sin\theta\right\rangle $
is obtained by multiplying the vorticity equation by $\sin\theta$
and then taking the flux surface average: 
\begin{eqnarray}
 &  & \pderiv{}{t}\nabla_{r}^{2}\left\langle \phi\sin\theta\right\rangle -\frac{\varepsilon_{n}}{2q}\nabla_{r}^{2}\left\langle A_{||}\cos\theta\right\rangle +\frac{\varepsilon_{n}}{2}\nabla_{r}\left[\left(1+\tau_{i}\right)\left\langle n\right\rangle +\tau_{i}\left\langle T_{i}\right\rangle +\tau_{i}\left\langle T_{e}\right\rangle \right]\nonumber \\
 &  & =-\left\langle \vec{\nabla}\cdot\left[\phi+p_{i},\vec{\nabla}_{\perp}\phi\right]\sin\theta\right\rangle +\frac{\beta}{2}\left\langle \left[A_{||},\nabla_{\perp}^{2}A_{||}\right]\sin\theta\right\rangle \label{smvor}
\end{eqnarray}
 The above equations are complemented by the transport like equations for
$\left\langle n_{e}\right\rangle $ and $\left\langle T_{i,e}\right\rangle $
\begin{eqnarray}
 &  & \pderiv{}{t}\left\langle n\right\rangle -\varepsilon_{n}\frac{\partial}{\partial r}\left[\langle\phi\sin\theta\rangle-\langle n_{e}\sin\theta\rangle-\langle T_{e}\sin\theta\rangle\right]=\langle\left[\phi,n_{e}\right]\rangle-\frac{\beta}{2}\langle\left[A_{||},J_{||}-v_{||}\right]\rangle+\left\langle S_{n}\right\rangle \label{mn}
\end{eqnarray}
 
\begin{eqnarray}
\frac{\partial}{\partial t}(\langle T_{i}\rangle-\frac{2}{3}\left\langle n\right\rangle )-\frac{5\varepsilon_{n}}{3}\nabla_{r}\left\langle T_{i}\sin\theta\right\rangle =-\left\langle \left[\phi,T_{i}-\frac{2}{3}n\right]\right\rangle +\left\langle S_{T_{i}}\right\rangle \label{mt}
\end{eqnarray}
 
\begin{eqnarray}
\frac{\partial}{\partial t}(\langle T_{e}\rangle-\frac{2}{3}\left\langle n\right\rangle )+\frac{5\varepsilon_{n}}{3}\nabla_{r}\left\langle T_{e}\sin\theta\right\rangle =-\left\langle \left[\phi,T_{e}-\frac{2}{3}n\right]\right\rangle +\left\langle S_{T_{e}}\right\rangle \label{mt-1}
\end{eqnarray}
 
The equations(\ref{mn},\ref{mt}) and (\ref{mt-1}) differ from the usual transport equations by 
the presence of a curvature term. This is due to the fact that these 
equations are obtained on taking flux surface average of the drift reduced equations 
(\ref{ern},\ref{it}) and(\ref{ert}) respectively where as the regular transport equations 
are obtained by taking the flux surface average of the 
starting fluid equations (\ref{GrindEQ__1_}-\ref{GrindEQ__4_}). As a consequence   
$\left\langle n_{e}\right\rangle $ and $\left\langle T_{i,e}\right\rangle $ also vary in 
GAM time scale rather than on slower transport time scale. The electron density and 
temperature equations arise because of the finite beta extension which demands 
non-adiabatic electron response. In electrostatic case with adiabatic electron response one still needs to keep the equation 
for $\langle T_{i}\rangle $ to be consistent. Many of the previous papers 
except Ref.\cite{falchetto:2007} on GAM happened to miss this somehow. 
Retention of this equation leads to another 
lower frequency branch with frequency going to zero when $k_{r}\to0$
which is distinctly different from the standard GAM whose frequency 
remains non-zero when $k_{r}\to0$. 

In the above equations $S_{n}$, $S_{T}$ and $S_{v}$ represents
particle, heat and momentum sources respectively. This system of equations
can be used to describe the complete nonlinear dynamics of GAMs using
a reduced drift-Braginskii description including the effects of finite
$\beta$ and collisionality. It is evident from the above equations that 
the GAM does contain
$m=1$ electromagnetic component which is in contrast to the previous 
Refs\cite{zhou:2007,lwang:2011} but in line with
the Refs\cite{smolyakov:2008,bashir:2014}. However these calculations can not explain the 
dominance of $m=2$ 
electromagnetic perturbation as observed in some experiment and simulations
\cite{berk:2006,meijere:2014} as the above 
equations are terminated at $m=1$. Extension to $m=2$ and beyond is deffered for future. 
The linearized form of the above equations can be written as a linear vector 
equation for the GAM state vector $G$  
\begin{eqnarray}
\frac{\partial G }{\partial t} = M G
\label{cr}
\end{eqnarray}
where $M$ is a coupling matrix with elements depending on $k_{r}, \tau_{i}$, and $q $. The 
matrix $M$ is provided in the appendix by equation (\ref{matm}).
Upto $m=1$ the GAM state vector $G$ is made of 
\begin{eqnarray}
G = (\langle n\rangle, \langle \phi \rangle, \langle T_{i} \rangle, \langle  T_{e} \rangle,
\langle n \sin \theta \rangle, \langle \phi \sin \theta \rangle, 
\langle A_{||} \cos \theta \rangle, \langle v_{||} \cos \theta \rangle, 
\langle T_{i}\sin\theta \rangle, \langle T_{e}\sin\theta \rangle) 
\end{eqnarray}
and $M$ is a $10\times 10$ matrix. 
In the $\nu\to 0$ but $\beta \ne 0$ limit it is straightforward to see that 
$\langle T_{e} \sin\theta\rangle = 0$ and hence $\langle T_{e} \rangle = \langle n\rangle$.
Hence the GAM state vector will become
\begin{eqnarray}
G = (\langle n\rangle, \langle \phi \rangle, \langle T_{i} \rangle,
\langle n \sin \theta \rangle, \langle \phi \sin \theta \rangle, 
\langle A_{||} \cos \theta \rangle, \langle v_{||} \cos \theta \rangle, 
\langle T_{i}\sin\theta \rangle ) 
\end{eqnarray}
and $M$ becomes a $8 \times 8$ matrix.
In the limit $\nu \to 0$ and $\beta \to 0$ from the above equations it follows that 
$\langle n \sin \theta \rangle = \langle \phi \sin \theta \rangle$, 
$\langle A_{||} \cos \theta \rangle = \langle T_{e} \sin \theta \rangle = \langle n \rangle = \langle T_{e} \rangle = 0$. Hence the 
GAM state vector reduces to 
\begin{eqnarray}
G = (\langle \phi \rangle, \langle T_{i} \rangle,
\langle n \sin \theta \rangle, \langle v_{||} \cos \theta \rangle, 
\langle T_{i}\sin\theta \rangle ) 
\end{eqnarray}
and $M$ becomes a $5\times 5 $ matrix.

\subsection{Linear GAM dispersion \label{sub:Linear-GAMs-dispersion}}

Linearizing the above set of equations (\ref{mvor}-\ref{smvor}), taking
Fourier transforms and neglecting $\langle n\rangle$, $\langle T_{i,e}\rangle$
, one obtains the following dispersion relation

\begin{eqnarray}
 &  & \omega^{2}\left[1-\frac{k_{r}^{2}(\varepsilon_{n}/2q)^{2}\left(2.14\omega+i(1.07/\nu)(\varepsilon_{n}/2q)^{2}\right)}{\left((\beta\omega/2+i\nu k_{r}^{2})\omega-(\varepsilon_{n}/2q)^{2}\right)\left(\omega+i(1.07/\nu)(\varepsilon_{n}/2q)^{2}\right)-1.21k_{r}^{2}\omega(\varepsilon_{n}/2q)^{2}}\right]\nonumber \\
 &  & =\left[\frac{\varepsilon_{n}^{2}}{2}+\left(\frac{\varepsilon_{n}}{2q}\right)^{2}\right]\left[1+\frac{5\tau_{i}}{3}\right.\nonumber \\
 &  & \left.+\frac{(2\omega/3)\left((\beta\omega/2+i\nu k_{r}^{2})\omega-(\varepsilon_{n}/2q)^{2}\right)+0.71k_{r}^{2}\omega(\varepsilon_{n}/2q)^{2}}{\left((\beta\omega/2+i\nu k_{r}^{2})\omega-(\varepsilon_{n}/2q)^{2}\right)\left(\omega+i(1.07/\nu)(\varepsilon_{n}/2q)^{2}\right)-1.21k_{r}^{2}\omega(\varepsilon_{n}/2q)^{2}}\right]\label{disp}
\label{freq0}
 \end{eqnarray}

In the limit $\nu\to0$ and $\beta\to0$ the above GAM dispersion
relation becomes: 
\begin{eqnarray}
\omega^{2}\left(1+k_{r}^{2}\right)=\left(\frac{\varepsilon_{n}^{2}}{2}
+\left(\frac{\varepsilon_{n}}{2q}\right)^{2}\right)\left(1+\frac{5\tau_{i}}{3}\right)
\label{freq}
\end{eqnarray}
consistent with the basic GAM frequency as obtained by many authors
(e.g. Ref \citep{hallatschek:2001,falchetto:2007}). However it is
slightly different in the temperature ratio dependence as obtained
fom the gyrokinetic calculations\citep{sugama:2006,gao:2006,gao:2008}
due to anisotropic temperature perturbations. 

\begin{figure}
\includegraphics[width=12cm,height=8cm]{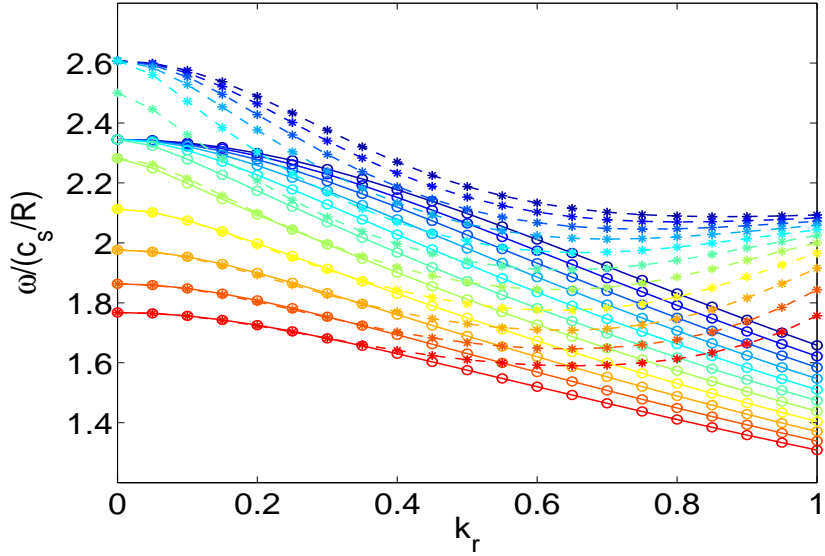} 
\protect\caption{GAM frequencies vs $k_{r}$ with $\beta$ and $\nu$ as parameters.
Color coding: blue$(\beta = 0)$ to red$(\beta = 0.04)$ in steps of
$0.004$. Line styles: solid for $\nu=0$ and dashed for $\nu=0.1$.
$\tau=1$, $q=4$. }
\label{wkr}
\end{figure}

In contrast \eqref{disp} includes the effects of finite $\beta$
and finite collisionality. Notice that these effects appear together
with $k_{r}$ in the above dispersion relation, implying that for
$k_{r}=0$, they can actually be neglected. We now explore some of
the charecteristics of the GAM frequency, especially its scalings
with $\beta$, $\nu$ and $k_{r}$ via the numerical solution of the
dispersion relation \eqref{disp}. Fig.\ref{wkr} shows the dispersion
properties of GAM. Without collisionalities the frequency decreases
with $k_{r}$ monotonically and frequency also decreases with $\beta$
at any $k_{r}$. This behavior is also consistent with the gyrokinetic
calculation in Ref.\citep{lwang:2011,bashir:2014}. At finite collisionality,
$\nu=0.1$, the GAM frequency shifts up. The amount of upshift depends
on $k_{r}$ at any given $\beta$. At low values of $\beta$ the upshift
in frequency is more towards $k_{r}\to0$ and $k_{r}\to1$ then when
$k_{r}\to0.5$. Whereas at higher $\beta$ values $(>0.02)$ the upshift
in GAM frequency is noticable only beyond $k_{r}=0.4$.

The radial group velocity at any $\beta$ and $k_{r}$ is either zero
or negative when $\nu=0$. Whereas at finite $\nu$ the radial group
velocity may either be negative, zero or positive depending on the
values of $\beta$, $\nu$ and $k_{r}$. 

\begin{figure}
\includegraphics[width=12cm,height=8cm]{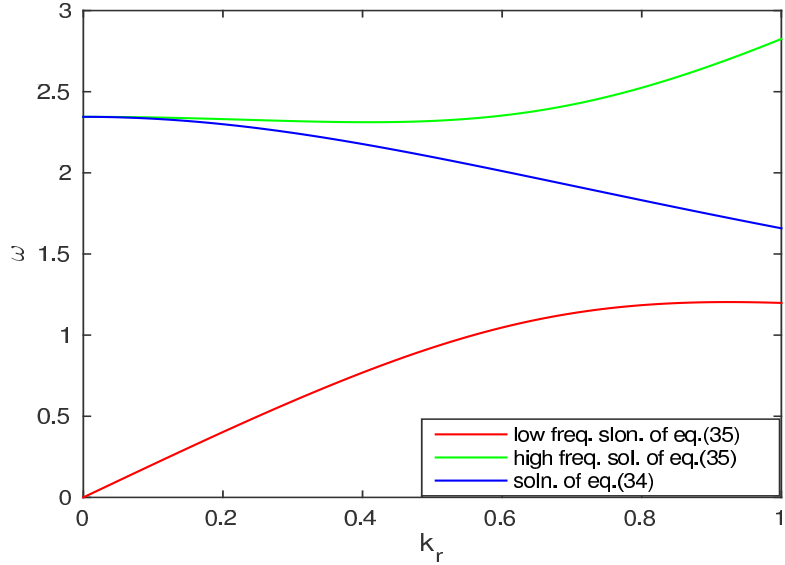} 
\protect\caption{GAM frequencies vs $k_{r}$ with effect of $\langle T_{i} \rangle$ for 
$\nu = 0$, $\beta = 0$}
\label{fig:new1}
\end{figure}

Self consistent inclusion of flux surface averaged temperature leads to the 
following modification of GAM dispersion in collisionless electrostatic limit 
\begin{eqnarray}
 \omega^{2}\left(1+k_{r}^{2}\right)= R_{Tn\sin\theta} + \left(\frac{\varepsilon_{n}^{2}}{2}
 +\left(\frac{\varepsilon_{n}}{2q}\right)^{2}\right)\left(1+ \tau_{i} + 
 \tau_{i}R_{Tsin\theta n\sin\theta}  \right)\label{freq1}
\end{eqnarray}
where the response $R_{Tsin\theta n\sin\theta}$ of $\langle T \sin\theta \rangle$ to $\langle n \sin\theta \rangle$ 
\[ 
 R_{Tsin\theta n\sin\theta} = \frac{2/3}{1 - \frac{1}{2} R_{TT\sin\theta}^{2}}
\]
and the response $R_{Tn\sin\theta}$ of $\langle T \rangle$ to $\langle n \sin\theta \rangle$
is given by 
\[ 
 R_{Tn\sin\theta} = R_{TT\sin\theta} R_{Tsin\theta n\sin\theta} 
\]
where $R_{TT\sin\theta}$ represents the response of $\langle T \rangle$ 
to $\langle T \sin\theta \rangle$
\[ 
 R_{TT\sin\theta} = -\frac{1}{\omega} \frac{5 \epsilon_{n}}{3}k_{r} 
\]
On switching off the response $R_{TT\sin\theta}$, the equation \ref{freq1} reduces to the 
simpler equation \ref{freq}. Solutions of equation (\ref{freq1}) are compared with the 
solutions of equation(\ref{freq}). It is seen that with self consistent 
treatment of the flux surface averaged ion temperature equation the frequency becomes 
non-monotonous in $k_r$; decreasing at small $k_r$ but increasing at large $k_r$. This can 
yield both inward and outward group velocites depending on the valaues of $k_r$. Moreover
another lower frequency branch appears (red curve in Fig.(\ref{fig:new1})) which 
increases with $k_r$ with the vanishing frequency at $k_{r} = 0$.  

\begin{figure}
\includegraphics[width=12cm,height=8cm]{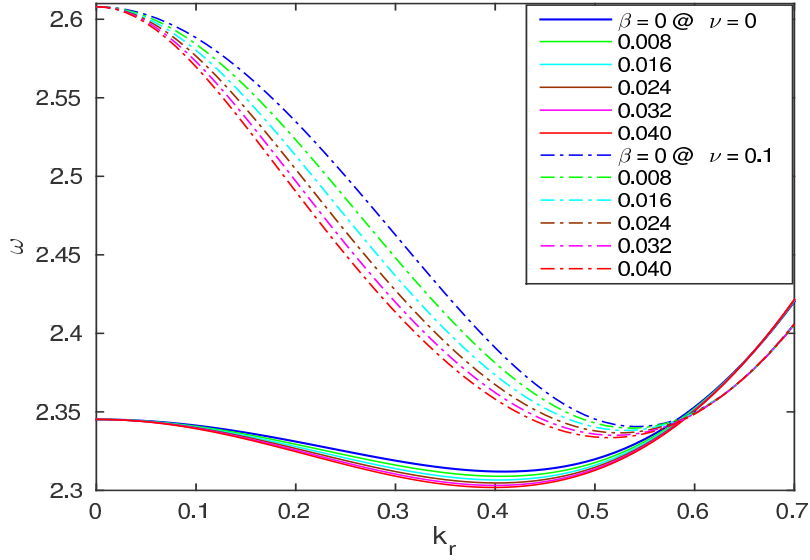} 
\protect\caption{GAM frequencies vs $k_{r}$ with effect of $\langle T_{i} \rangle$ 
and  $\langle n \rangle$ for $\nu = 0 $ and $0.1$ for different values of  $\beta$}
\label{fig:new2}
\end{figure}

Inclusion of $\langle T \rangle$ and $\langle n \rangle$
responses in the more general dispersion relation equation \ref{freq0} for finite 
collisionality and beta is straightfoward but tedious. The most general dispersion relation 
in principle can be obtained from 
\begin{eqnarray}
 |\lambda I - M| = 0
 \label{freq2}
\end{eqnarray}
where the imaginary part of the eigenvalue $\lambda$ gives the real frequency and the 
real part of $\lambda$ gives the growth rate. It can best be studied 
numericaly for the eigenvalues $\lambda$ of the matrix $M$. Fig.\ref{fig:new2} 
shows the dispersion relation with the self-consistent treatment of the average density and 
temperature equations. It is seen that in general the frequency decreases with $\beta$ 
with the rate of decrese depending on the values of $k_r$ and $\nu$. However for a 
given $\beta$ value the frequency increases with $\nu$ for $0 \le k_{r}< 0.58$ and 
decreases with $\nu$ for $k_{r} \gtrsim 0.58$. The major differece between the exact 
dispersion relations Fig.\ref{fig:new2} and the approximate one in Fig.\ref{wkr} are the 
following. The collisionality may enhance or reduce GAM frequency depending on the value 
of $k_{r}$ in the exact case whereas collisionality has always up shifting effect on GAM 
frequency in the approximate model. Also in the exact model the down shifting effect of 
$\beta$ on GAM frequency is much less pronounced than in the approximate model.    

\subsection{Comparison with experiment\label{sub:Comparison-with-experiment}}

\begin{figure}[h!]
\centering{}\includegraphics[width=14cm,height=7cm]{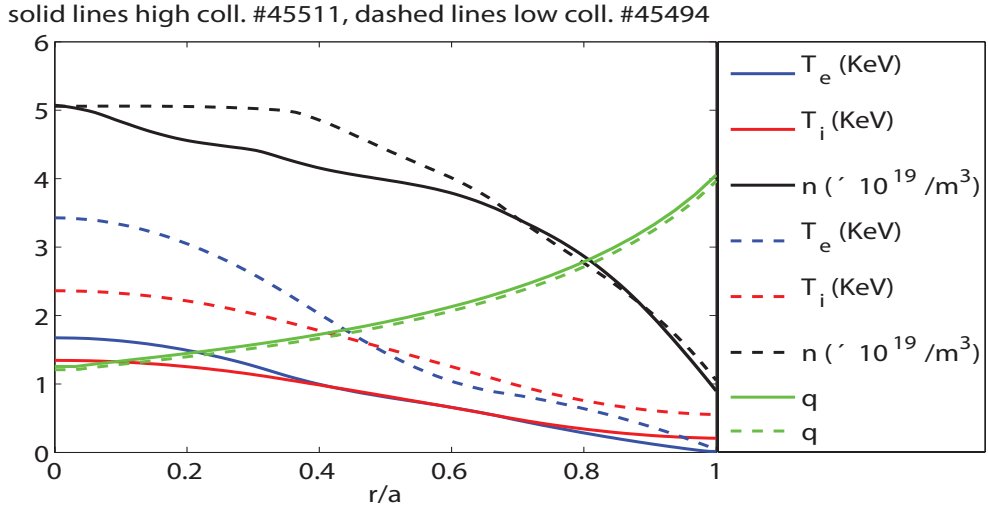} 
\protect\caption{Equilibrium profiles at two different collisionalities. Solid lines
are from high collisionality shot $\#45511$and dashed lines from
low collisionality shot $\#45494$.}
\label{fig:profiles}
\end{figure}

\begin{figure}[h!]
\centering{}\includegraphics[width=7cm,height=7cm]{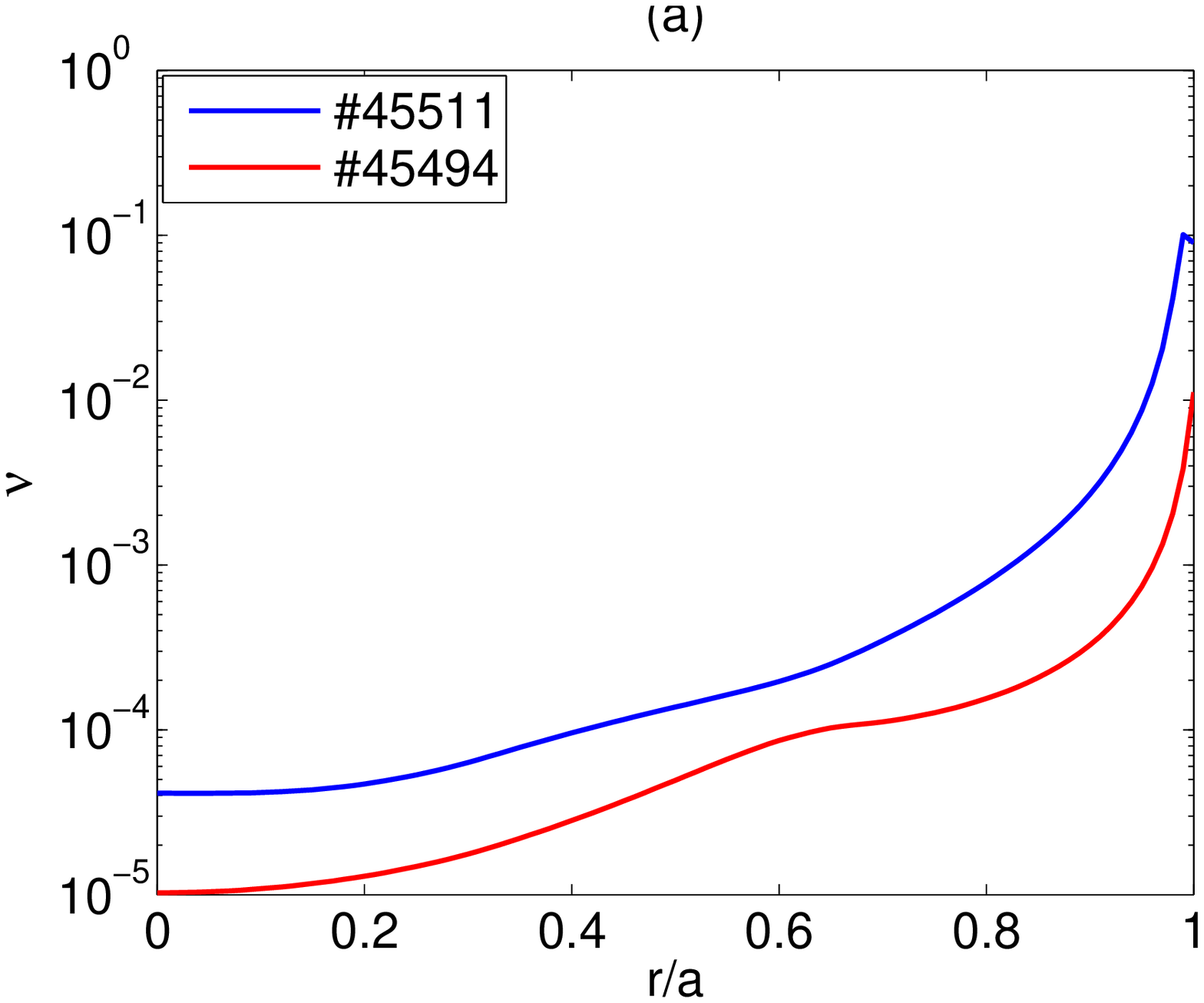} 
\includegraphics[width=7cm,height=7cm]{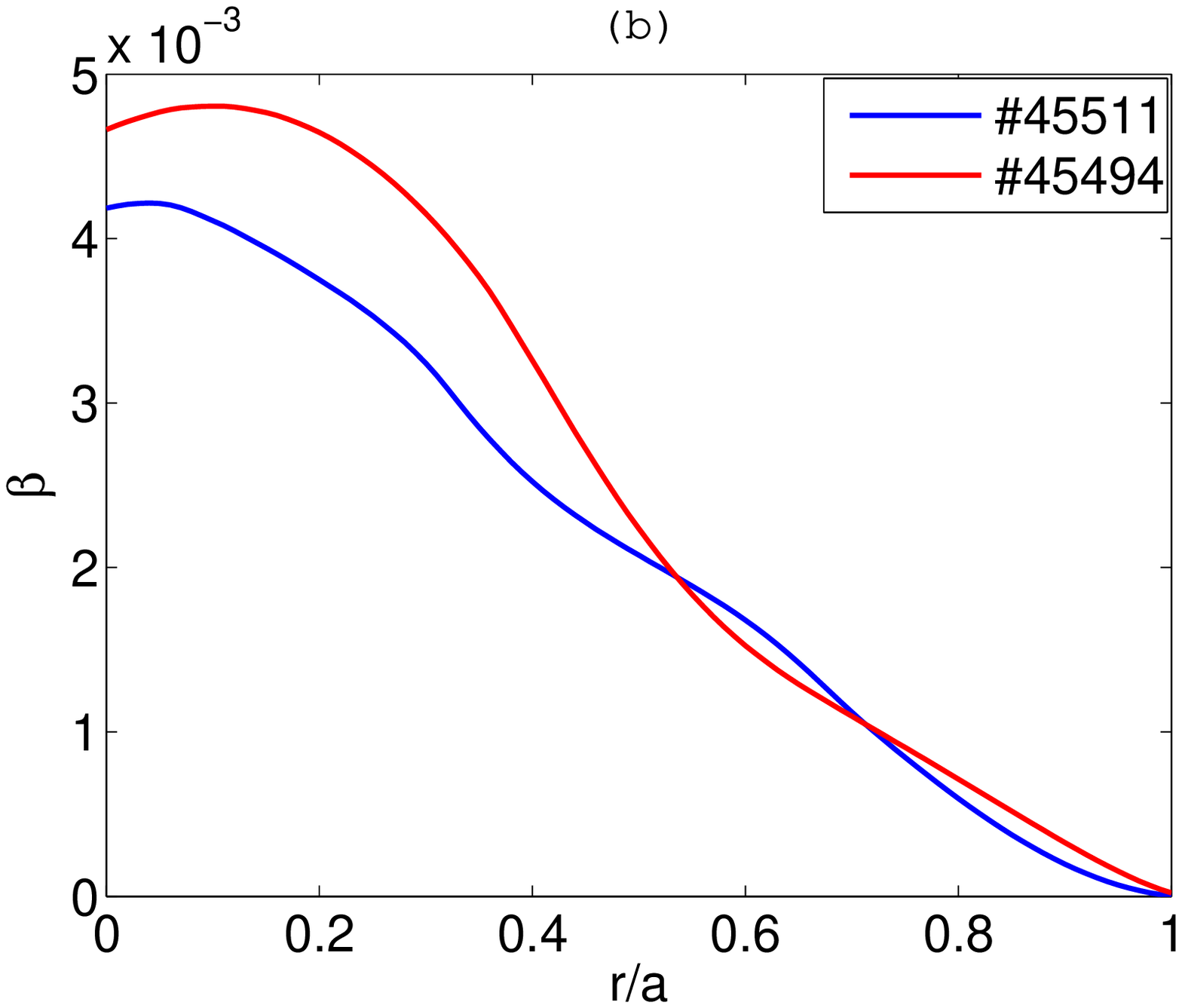}
\protect\caption{(a) Radial profiles of collisionalities, (b) Radial profiles of plasma
beta. }
\label{fig:fig3}
\end{figure}

We compare the theoretical GAM frequency as given by the dispersion
relations \eqref{disp}, \eqref{freq2} and the with the 
experimental GAM frequency observed
in the Tore Supra tokamak for two different values of collisionality
(i.e. shots $\#45494$ and $\#45511$) \citep*{vermare:2011}. Equilibrium
profiles of density and temperature are shown in Fig.\ref{fig:profiles}.
The temperature ratio $\tau_{i}$ is greater than $1$ towards the
edge but less than $1$ towards the core. The density remains almost
the same towards the edge in these two discharges. Radial profiles
of collisionality and beta, which are calculated using these equilibrium
profiles, are shown in Fig.\ref{fig:fig3}.

\begin{figure}[h!]
\centering{}\includegraphics[width=16cm,height=7cm]{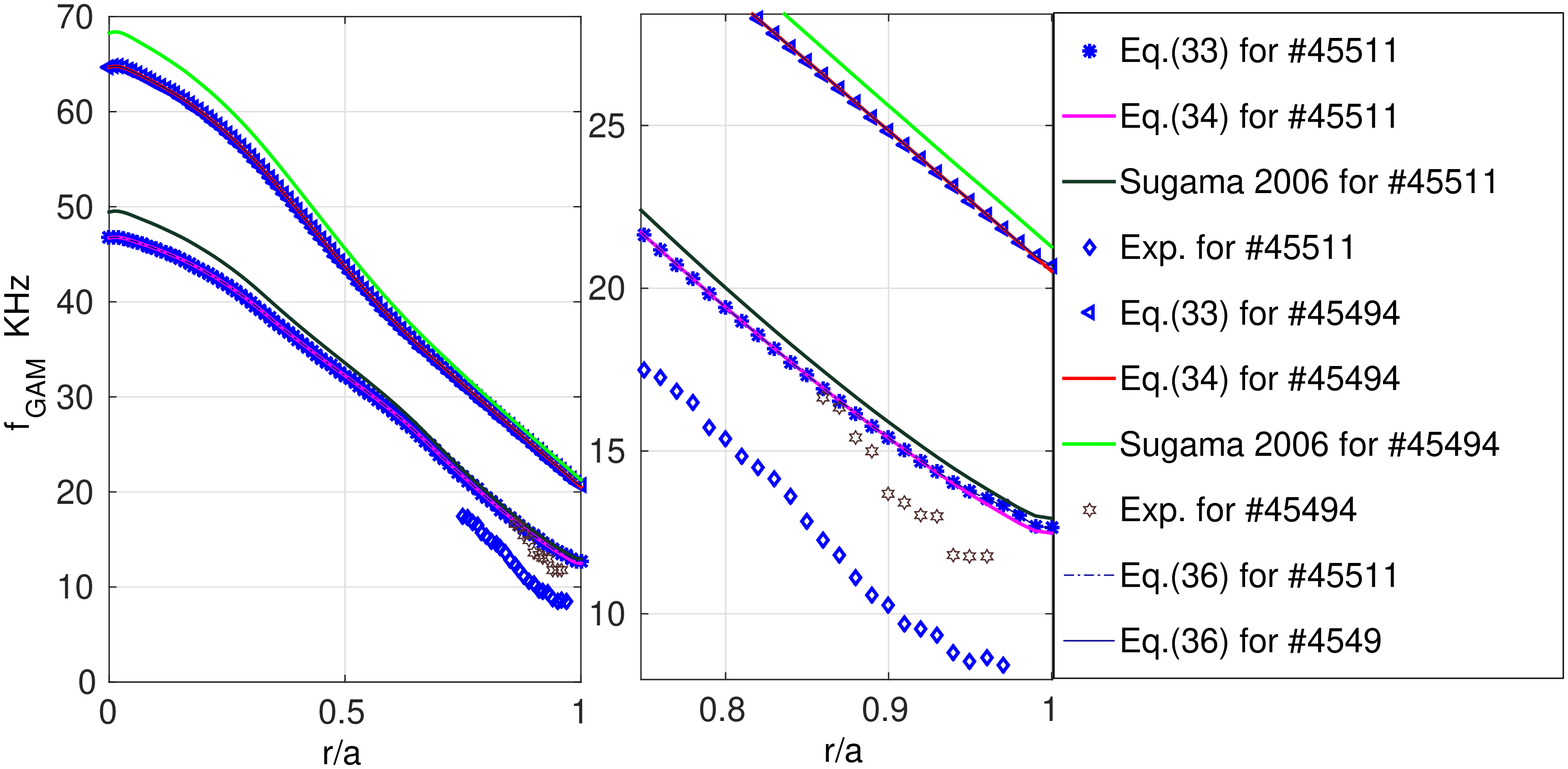} 
\protect\caption{Comparisons between experiment and theory.}
\label{fig:fig4}
\end{figure}

The radial profiles of experimental GAM frequencies are compared against
the theoretical values as obtained from equations \eqref{disp},
\eqref{freq}, \eqref{freq2} and from Sugama's formula\citep{sugama:2006} for $k_{r}=0$
are shown in Fig. \ref{fig:fig4}. Notice that our theoretical frequency
compares well with Sugama's for both the shots but experimental frequencies
are lower than both. The frequency increases inward from the edge,
which is consistent with increase of temperature, but the absolute
values of experimental frequencies are about $50\%$ below the theoretical
values, consistently. We discuss below if this is due to finite $k_{r}$.
Moreover, the experimental frequency is higher in low collisionality
shots than in high collisionality shots. This might give the impression
that GAM frequency goes down with collisionality, which is opposite
to the theoretical prediction shown in the Fig.\ref{wkr}? This is
in fact probably due to the change in temperature profiles between
the two shots, rather than the effect of collisionality.

The profiles of theoretical GAM frequencies with and without collisions
differs only slightly towards the edge where collisionality is higher
than in the core. This shows that the collisionality and beta values
are too low to make an appreciable impact on the frequency. However
the absolute values of GAM frequencies are higher for low collisionality
shots than for high collisionality shots, which is consistent with
the upshift in temperature profiles as seen in Fig.\ref{fig:profiles}. 

\begin{figure}[h!]
\centering{}\includegraphics[width=14cm,height=7cm]{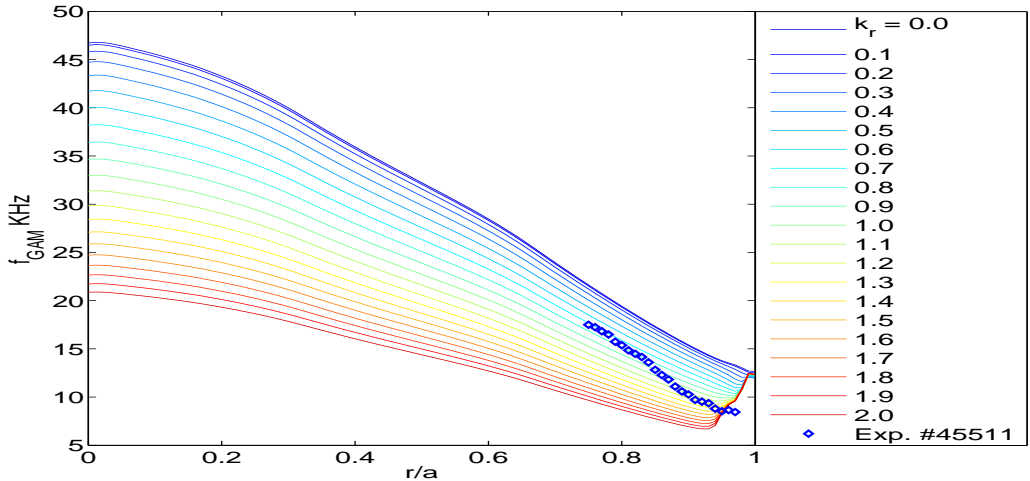} 
\protect\caption{Radial scan of GAM frequency with $k_{r}\rho_{s}$ as parameter for
high coll. shots $\#45511$}
\label{fig:fig5}
\end{figure}

\begin{figure}[h!]
\centering{}\includegraphics[width=14cm,height=7cm]{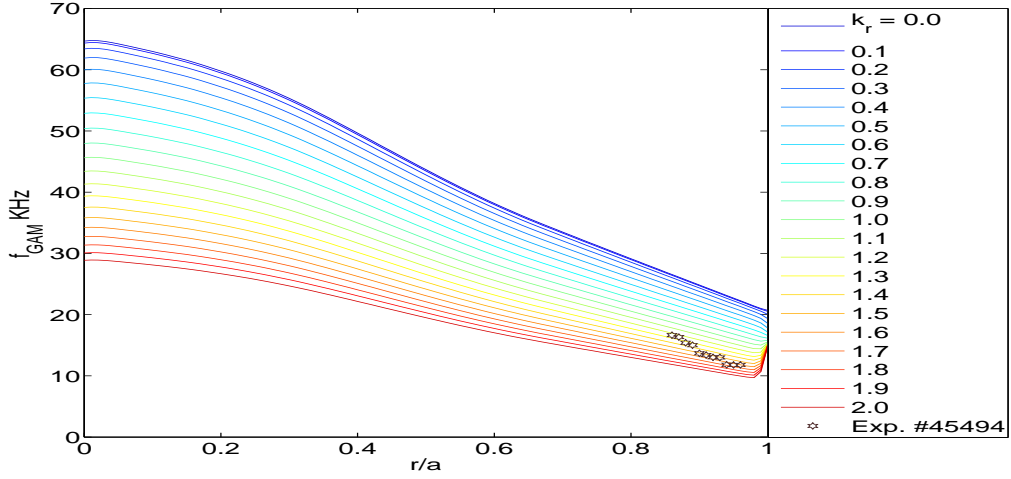}
\protect\caption{Radial scan of GAM frequency with $k_{r}\rho_{s}$ as parameter for
low coll. shots $\#45494$}
\label{fig:fig6}
\end{figure}

If one tries to explain the observed discrepancy between the theoretical
and the experimental frequencies by invoking a finite $k_{r}$ for
the GAM one notes that the GAM frequency, as predicted by Eqn \ref{disp}
goes down with increasing $k_{r}$ at any radius as shown in Figs.\ref{fig:fig5}
and \ref{fig:fig6}. However it goes up with $k_{r}$ at the edge
since collisionality is felt stronger at shorter scale lengths. The
frequency values computed by assuming a finite $k_{r}$ intersect
with experimental observations for different wave numbers at different
radii. In order to match with experimental observations, the radial
wavenumber of the GAM has to increase with radius. The overlapping
region of $k_{r}$ for high collisionality shots is $[0.7,1.5]$ and
for low collisionality shots is $[1.3,1.7]$.

\begin{figure}[h!]
\centering{}\includegraphics[width=14cm,height=7cm]{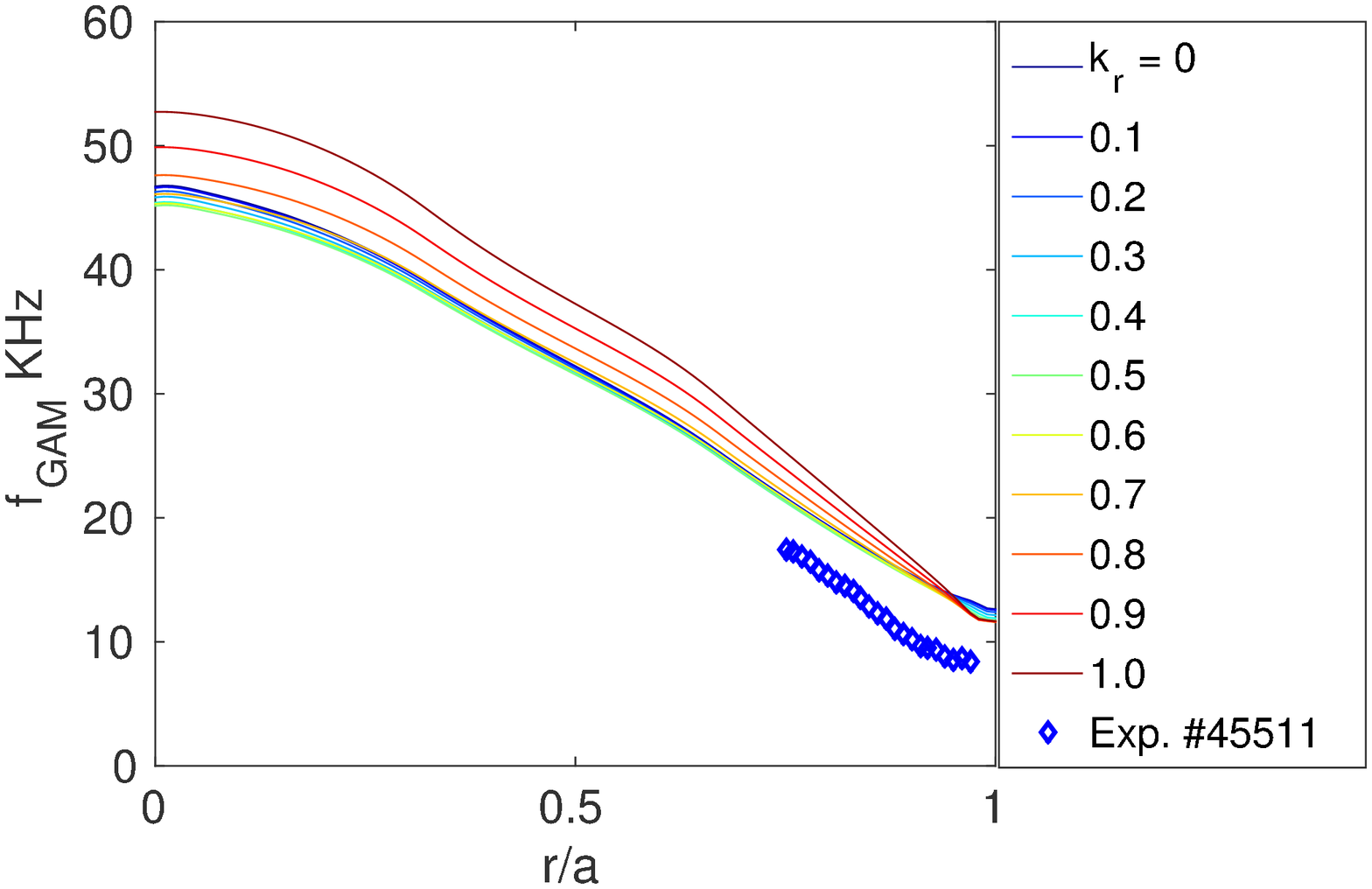}
\protect\caption{Radial scan of GAM frequency with $k_{r}\rho_{s}$ as parameter for
high coll. shots $\#45511$, with self-consist treatment of $\langle T_{i,e} \rangle$ and
$\langle n \rangle$}
\label{fig:new3}
\end{figure}

\begin{figure}[h!]
\centering{}\includegraphics[width=14cm,height=7cm]{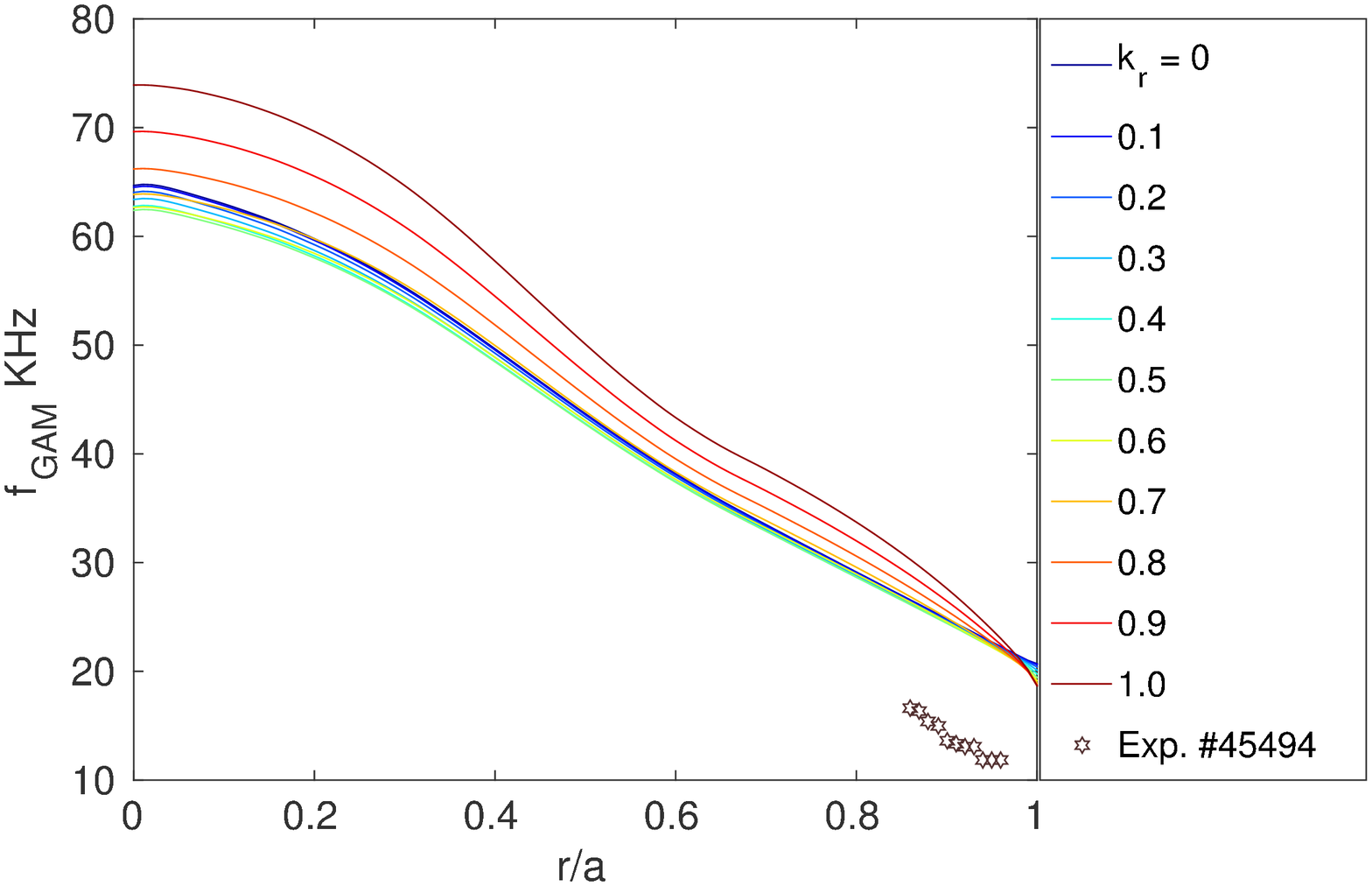}
\protect\caption{Radial scan of GAM frequency with $k_{r}\rho_{s}$ as parameter for
low coll. shots $\#45494$, with self-consist treatment of $\langle T_{i,e} \rangle$ and
$\langle n \rangle$  }
\label{fig:new4}
\end{figure}

However we think that such a profile of $k_{r}$ is rather unrealistic
and is probably not the explanation of the observed discrepancy. The reason 
being that on self-consistent inclusion of $\langle n \rangle$ and 
$\langle T_{i,e} \rangle$ responses through equation \eqref{freq2} breaks the 
monotonically decreasing behavior of GAM frequency on $k_{r}$.  Another 
important reason for this conclusion is that, in fact when other harmonics
of the GAMs are considered (i.e. $m=2$, $m=3$ etc.), which are linearly
coupled to $m=1$, the $k_{r}$ dependence may be observed to change
substantially. The gap between experiment and theory increases with $k_{r}$
at high $k_{r}$ on self-consistent inclusion of $\langle n \rangle$ and 
$\langle T_{i,e} \rangle$ responses at any radius as can be seen in the figures 
(\ref{fig:new3}) and (\ref{fig:new4}).

\section{Conclusions\label{sec:Conclusions}}

Using the reduced Braginskii equations under the drift approximation,
a set of nonlinear electromagnetic equations retaining plasma beta
and electron ion collisionality were obtained. Appropriate flux surface
averaging were applied on the resulting set of equations in order
to derive a fully nonlinear set of equations for the GAMs. This approach
clearly shows that the GAM perturbations consist of $\langle\phi\rangle$,
$\langle\phi\sin\theta\rangle$, $\langle n\sin\theta\rangle$, $\langle T_{i,e}\sin\theta\rangle$,
$\langle v_{\parallel}\cos\theta\rangle$,$\langle A_{\parallel}\cos\theta\rangle$
up to the first poloidal harmonic. These equations can be used for
studying both the nonlinear drive and the linear oscillation of the
GAM in a collisional, electromagnetic model of the plasma edge. However,
one needs a better closure\citep{hammet:1990,Beer:1995,sugama:2001}
 in order to describe the collisionless GAM damping, which can in
principle be included in the current model. Note however that initial
studies in Tore Supra suggest that GAM damping near the edge region
is mainly collisional as well.

Linearizing these equations, a general linear dispersion relation
is obtained, which contains the effects of electron ion collisionality
and plasma beta. Following results are obtained without $\langle n \rangle$ and 
$\langle T_{i,e} \rangle$. At zero collisionality and beta, the GAM frequency
monotonically decreases with $k_{r}$. At finite beta the GAM frequency
shifts down preserving the monotonically decreasing nature with $k_{r}$.
At finite collisionality and low beta the frequency is shifted up,
decreasing in $k_{r}$ at low $k_{r}$ and increasing at high $k_{r}$.
The upshift in frequency with collisionality is more prominent at
low and high $k_{r}$ at low beta but, at high beta the collisional
upshift is seen only at high $k_{r}$ values. However with self-consistent 
responses of $\langle n \rangle$ and $\langle T_{i,e} \rangle$ the results are modified 
to the following. At zero collisionality and beta, the GAM frequency
decreases with $k_{r}$ at low $k_{r}$ and incerases at high $k_{r}$. At finite beta the 
GAM frequency
shifts down preserving the same non-monotonic decreasing nature. However the amount of 
down shift is more prominent in a small wave number range centered around $k_{r}\sim 0.4$.
At finite collisionality the frequency is shifted up at low $k_{r}$ and shifted down at 
high $k_{r}$, still preserving the non-monotonic behavior of frequency 
decreasing in $k_{r}$ at low $k_{r}$ and increasing at high $k_{r}$.
The upshift in frequency with collisionality is more at
low $k_{r}$ than the down shift at high $k_{r}$.
We argue that these
linear trends for the GAM frequency should be retained in the nonlinear
regime, since the nonlinearity acts mainly as a source term for the
GAM, and does not modify its response.

The theoretical GAM frequencies obtained this way, were then compared
to two TORE SUPRA shots, which were part of a collisionality scan
and a consistent discrepancy were observed where the absolute values
of experimental frequencies were about $50\%$ below the theoretical
values for $k_{r} = 0$. In fact, without the 
$\langle n \rangle$ and 
$\langle T_{i,e} \rangle$ responses 
the experimental GAM frequencies 
may be tailored
to match the theoretical values by assuming finite $k_{r}$ values
in the range $k_{r}\in[0.7,1.5]$ for the high collisionality shots
and $k_{r}\in[1.3,1.7]$ for the low collisionality shot. However
since these values are rather large, and since the $k_{r}$ dependence
is mainly a feature of taking only the first poloidal harmonic of
GAM, it is argued that ``finite $k_{r}$ effects'' probably does
not explain the observed discrepancy. Also the self consistent accounting of 
$\langle n \rangle$ and 
$\langle T_{i,e} \rangle$ responses makes the experiment-theory disagreement even worst on
increasing $k_{r}$.

Note also that while the experimental GAM frequencies are lower at
high collisionality than at low collisionality, this observed trend
is due to the change in the temperature profiles and not
due to the change in collisionality and plasma beta. This implies
that in order to scan the collisionality dependence of the GAM frequency,
one has to keep the temperature constant. Since these shots were part
of a collisionality scan for the confinement, it was the other dimensionless
variables, such as $\rho_{*}$ etc. who were kept constant.

These results leave the question of the discrepancy between the GAM
frequency measured in Tore Supra and the theoretical prodictions.
Notice that the discrepancy is equally important if one uses a more
complex gyrokinetic formula, which is shown in figure \ref{fig:fig4}.
We think that future work should include higher GAM harmonics in a
similar fluid model of the edge, which may hopefully resolve this
discrepancy.
\begin{acknowledgments}
The authors are thankful to X Garbet, Y Sarazin, and G Dif-Paradalier,
T.S. Hahm and P.H. Diamond for discussions. This work has been 
supported by funding from the "Departement
D'Enseignement-Recherche de Physique Ecole Polytechnique" and it ihas been carried out within the framework of the EUROfusion Consortium and has received
funding from the European Union's Horizon $2020$ research and innovation programme under grant
agreement number $633053$. This work was also
partly supported by the french ``Agence
National de la Recherche'' contract ANR JCJC 0403 01. The views and opinions expressed herein do not necessarily reflect those
of the European Commission. The authors are also 
extremley thankful to the annonymous referee for careful refereing and the suggestions for 
the improvement of the paper.
\end{acknowledgments}
\appendix
\section{The matrix M}

The matrix $M$ in the equation \ref{cr} is 
\begin{eqnarray}
M = (M_{n}; M_{\phi}; M_{T_{i}}; M_{T_{e}}; M_{n\sin\theta}; 
M_{\phi\sin\theta}; M_{A_{||}\cos\theta}; M_{v_{||}\cos\theta}; M_{T_{i}\sin\theta}; 
M_{T_{e} \sin\theta})
\label{matm}
\end{eqnarray}
where 
\begin{eqnarray}
 && M_{n} = (0,0,0,0,-i\epsilon_{n} k_{r},  i\epsilon_{n} k_{r}, 0,0,0, i\epsilon_{n} k_{r})\\
 && M_{\phi} = (0,0,0,0, i\epsilon_{n} (1+\tau_{i})/ k_{r}, 0,0,0, i\epsilon_{n} \tau_{i}/ k_{r}, 
 i\epsilon_{n}/ k_{r})\\
&& M_{T_{i}} = (0,0,0,0, -i 2\epsilon_{n}  k_{r}/3, i2\epsilon_{n} k_{r}/3, 0,0, 
 i5\epsilon_{n} k_{r}/3, - i 2\epsilon_{n}  k_{r}/3)\\
&& M_{T_{e}} = (0,0,0,0, -i 2\epsilon_{n}  k_{r}/3, i2\epsilon_{n} k_{r}/3, 0,0,0, 
- i7\epsilon_{n} k_{r}/3 )\\
&& M_{n\sin\theta} = (-i \epsilon_{n}  k_{r}/2, i \epsilon_{n}  k_{r}/2, 0, 
 -i \epsilon_{n}  k_{r}/2,0,0, -\epsilon_{n} k_{r}^{2}/2q, \epsilon_{n}/2q, 0,0)\\
&& M_{\phi\sin\theta} = (-i \epsilon_{n}(1+\tau_{i})/2 k_{r}, 0, i \epsilon_{n} \tau_{i}/2 k_{r},
i \epsilon_{n}/ 2k_{r}, 0,0, \epsilon_{n} /2q, 0, 0,0)\\
&& M_{A_{||}\cos\theta} = (0,0,0,0, \epsilon_{n}/\beta q,  -\epsilon_{n}/\beta q , 
-\nu k_{r}^{2}2/\beta, 0,0, 1.71\epsilon_{n}/\beta q)\\
&& M_{v_{||}\cos\theta} = (0,0,0,0, -\epsilon_{n}(1+\tau_{i})/2 q, 0,0,0,  -\epsilon_{n} \tau_{i}/2q ,
-\epsilon_{n}/2q)\\
&& M_{T_{i}\sin\theta} = ( - i \epsilon_{n}k_{r}/3, i \epsilon_{n}k_{r}/3,  0, 
-i \epsilon_{n}k_{r}7/6, 0,0,  -\epsilon_{n} k_{r}^{2} 1.38/2 q, \epsilon_{n} /6q ,
0, -1.07 (\epsilon_{n}/2q)^{2}/\nu)
\end{eqnarray}

%

\end{document}